\documentclass[ floats, 12pt,doublespace ]{article}

\usepackage{graphicx}

\title{ 2D and 3D quantum rotors in a crystal field:
  critical points, metastability, and reentrance}
\author{Yuri A. Freiman,$^1$  Bal\'azs Het\'enyi,$^2$ and
    Sergei M.  Tretyak$^1$ \\ \\
$^1$B. Verkin Institute of Low Temperature Physics 
  and Engineering \\ 47 Lenin Avenue, 
  Kharkov, UA-61103, Ukraine \\ \\
$^2$Institut f\"ur Theoretische Physik \\ Technische Universit\"at Graz\\Petersgasse 16,  A-8010 Graz, Austria
}

\date{\today}

\begin{document}

%%%%%%%%%%%%%%%%%%%%%%%%%%%%%%%%%%%%%%%%%%%%%%%%%%%%%%
\maketitle
\begin{abstract}
  An overview of results of models of coupled quantum rotors is presented.
  We focus on rotors with dipolar and quadrupolar potentials in two and three
  dimensions, potentials which correspond to approximate descriptions of real
  molecules adsorbed on surfaces and in the solid phase.  Particular emphasis
  is placed on the anomalous reentrant phase transition which occurs in both
  two and three-dimensional systems.  The anomalous behaviour of the entropy,
  which accompanies the reentrant phase transition, is also analyzed and is
  shown to be present regardless if a phase transition is present or not.
  Finally, the effects of the crystal field on the phase diagrams are also
  investigated.  In two-dimensions the crystal field causes the disappearance
  of the phase transition, and ordering takes place via a continuous increase
  in the value of the order parameter.  This is also true in three dimensions
  for the dipolar potential.  For the quadrupolar potential in three
  dimensions turning on the crystal field leads to the appearance of critical
  points where the phase transition ceases, and ordering occurs via a
  continuous increase in the order parameter.  As the crystal field is
  increased the range of the coupling constant over which metastable states
  are found decreases.
\end{abstract}
%% end of \twocolumn
%%%%%%%%%%%%%%%%%%%%%%%%%%%%%%%%%%%%%%%%%%%%%%%%%%%%%%

\section{Introduction}

\label{sec:intro}

In molecular solids the energy scales of translation, rotation, and vibration
can be expected to be of different orders of magnitude.  In the solid
hydrogens~\cite{Silvera80,Kranendonk83,Mao94}, the rotational lines are
clearly distinct from the spectral signatures of the translations and
rotations.  For a large pressure interval in such systems models of coupled
rigid rotors are sufficient to understand the general features of phase
transitions in particular those of the orientational kind.  In this work we
refer to such systems as orientational crystals.

There exist systems in both two and three dimensions which can be thought of
as orientational crystals.  Two dimensional examples are physisorbed molecules
on inert surfaces, such as N$_2$ or H$_2$ and its isotopes on graphite or
boron-nitride.  The former can be approximated by a model of planar rotors,
known as the anisotropic planar rotor (APR)
model~\cite{OShea79,OShea82,Mouritsen82,Marx96}.  This model exhibits an
orientational order-disorder phase transition from an orientationally
disordered state to the orientationally ordered herringbone structure.  In
N$_2$ on a graphite surface this transition takes place at 30
K~\cite{Chung77,Eckert79}, well below the liquid-solid ordering temperature
of 47 K~\cite{Kjems74}.  While the classical APR model accounts for the
orientational ordering, a more quantitative description of the system
necessitates the inclusion of quantum effects~\cite{Presber98}.  Also, models
of coupled quantum planar rotors are useful in describing other systems, such
as granular
superconductors~\cite{McLean79,Simanek80,Maekawa81,Doniach81,Simanek85,Simkin91}
and more recently the bosonic Hubbard model~\cite{Polak08,Kopec04}.

Three-dimensional examples are the solid phases of the hydrogens and
different isotopes.  The behavior of the hydrogens is generally made more
complex by the fact that ortho-para conversion times are slow on the
time-scale of rotations (in the pressure ranges considered here
$\leq100$GPa)~\cite{Strzhemechny00}, hence it is a reasonable approximation
to take the ortho-para ratio to be a fixed parameter~\cite{Harris85}.  The
existence of ortho and para species is due to the coupling of nuclear spins
and the rotational quantum numbers characterizing a particular molecule.  For
the H$_2$ molecule a rotation of angle $\pi$ corresponds to an exchange of
the constituent atoms, hence the wavefunction has to be anti-symmetric in
such a rotation.  Since the H atoms are of spin $\frac{1}{2}$, the possible
spin states of the molecule as a whole are three symmetric and one
anti-symmetric spin state.  To preserve the overall antisymmetry of the
wavefunction the symmetric spin states couple with anti-symmetric spatial
states (odd angular momentum or odd-$J$) and the anti-symmetric spin states
couple with symmetric spatial states (even-$J$).  In HD, where the atoms are
indistinguishable all angular momentum states are allowed (all-$J$).  In
D$_2$ the constituent atoms are bosons, hence the wavefunction has to be
symmetric.  However this leads to a qualitatively similar situation: here
symmetric(anti-symmetric) spin-states are even-$J$(odd-$J$).

The orientational ordering properties of odd-$J$, even-$J$, and all-$J$
systems show striking differences.  Odd-$J$ systems show orientational
ordering in the ground state, whereas even-$J$ systems order at higher
pressures.  At low pressures and low temperatures the even-$J$ systems can be
thought of as spheres.  An interesting anomalous feature that was first
predicted for all-$J$ systems is the {\it reentrant} phase diagram.  Upon
cooling, in certain pressure ranges, the system orders orientationally due to
a decrease in thermal fluctuations.  In the reentrant region, the
orientationally ordered phase is destroyed by quantum fluctuations (also know
as quantum melting).  This effect was first predicted in the mean-field phase
diagram of the all-$J$ hydrogen system (HD)~\cite{Freiman91,Brodyanskii93},
and experimentally verified thereafter~\cite{Moshary93}.  

For the quantum generalization of the APR model (QAPR) the system
corresponding to the all$-J$ case also shows reentrance.  This was first
predicted by mean-field theory~\cite{Martonak97}, and then verified via
quantum Monte Carlo calculations~\cite{Muser98} as well as quantum Monte
Carlo calculations analyzed via finite size
scaling~\cite{Hetenyi99,Hetenyi01}.  Reentrance was also found in the
corresponding model of granular superconductors~\cite{Simanek85,Simkin91}.
\begin{figure}[htp]
\begin{center}
\vspace{1cm}
\includegraphics[width=7cm,height=5cm]{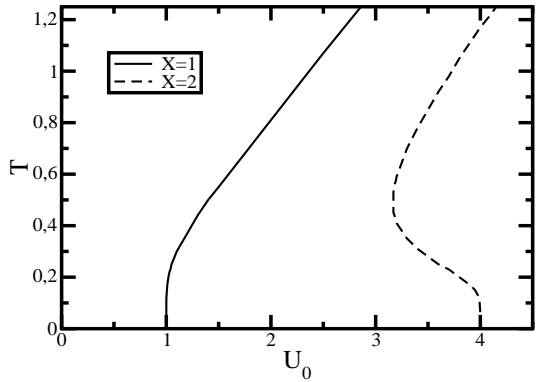}
\end{center}
\caption{Phase diagrams for the systems without crystal field for $X=1,2$.}
\label{fig:mf2dm1m2}
\end{figure}

Recently a set of studies~\cite{Hetenyi05a,Freiman05,Hetenyi05b} have suggested
that if the thermal equilibrium distribution of the ortho-para ratio is
reached then reentrance can occur in the homonuclear systems H$_2$ and D$_2$.
This conclusion is supported by experimental evidence~\cite{Goncharenko05}.

In this paper we give an overview of the mean-field theory of phase
transitions in coupled rotors with particular attention to the issues of
reentrance, other quantum anomalies, and meta-stability.  We comparatively
analyze coupled planar rotors (two-dimensional model) and coupled linear
rotors (three-dimensional).  We show that the dipolar potential does not
exhibit the reentrance anomaly, whereas the quadrupolar one does.  The phase
transition turns out to be second order in all cases except for the linear
rotors in a quadrupolar potential where it is first order.  We also
investigate the effects of the crystal field: in the case of the linear rotor
model with quadrupolar potentials the crystal field causes the appearance of
critical points which separate lines of the phase diagram where the
transition is first order from regions where there is no phase transition,
but simply a continuous change of the order parameter~\cite{Freiman03}.  We
show that the range over which meta-stable states (which accompany a
first-order phase transition) depends on the crystal field: as it is
increased this region becomes smaller, and disappears when the phase
transition itself disappears.  We also analyze the behaviour of the entropy
in all cases.

\section{Coupled rotors in two dimensions}
\label{sec:qapr}

The model we study in this section is described by the Hamiltonian
\begin{equation}
\label{eqn:H}
H = -B \sum_{i=1}^N \frac{\partial^2}{\partial \phi_i^2} - \frac{U}{2} \sum_{\langle
  i,j \rangle} \mbox{cos}(X\phi_i)\mbox{cos}(X\phi_j) 
   - U_1 \sum_{i=1}^N \mbox{cos}(X\phi_i),
\end{equation}
where $B$, $U$, and $U_1$ denote the rotational constant, the coupling
constant, and the strength of the crystal field respectively, and where the
sum $\langle i,j \rangle$ runs over nearest neighbors.  The parameter $m$
specifies the periodicity of the potential.  In this work we will investigate
the cases $X=1,2$, which show qualitatively different behaviour.  As a unit
of energy and temperature we choose the rotational constant $B$ in all of the
subsequent cases.

Applying the mean-field approximation to this Hamiltonian results in
\begin{equation}
\label{eqn:H_MF}
H_{MF} = -B \sum_{i=1}^N \frac{\partial^2}{\partial \phi_i^2} 
- (U_0\gamma + U_1) \sum_{i=1}^N \mbox{cos}(X\phi_i) 
+ \frac{U_0 N}{2}\gamma^2,
\nonumber
\end{equation} 
where $\gamma$ denotes the order parameter, and $U_0=Uz$ with $z$ denoting
the coordination number.  We note that had we used the
dipole-dipole (quadrupole-quadrupole) potential in Eq. (\ref{eqn:H}) the
resulting approximate Hamiltonian can be shown to be the same as the one in
Eq. (\ref{eqn:H_MF}) with $X=1$($X=2$) with a modified coupling constant.
\begin{figure}[htp]
\begin{center}
\vspace{1cm}
\includegraphics[width=7cm,height=5cm]{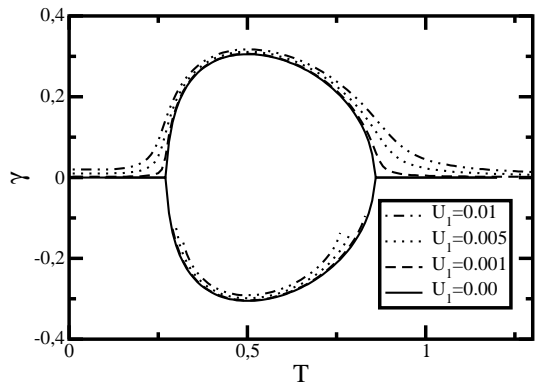}
\end{center}
\caption{Order parameter as a function of temperature for the $X=2$ system at
  $U_0=3.50$.  The curves with negative values indicate a metastable state in
  the case of finite crystal field.}
\label{fig:op2d}
\end{figure}

The mean-field phase diagrams without crystal field for $X=1,2$ are shown in
Figure \ref{fig:mf2dm1m2}.  The phase diagrams separate the orientationally
disordered phase (at lower values of the coupling constant) from the
orientationally ordered phase.  The two striking differences between the two
curves are the quantitative difference between the onset of order and the
shape of the phase diagram.  The former can be attributed to the width of the
barrier through which the quantum systems tunnel.  The $X=1$ system has a
wider barrier than the $X=2$ system.  The reentrance has been found in the
related QAPR model via quantum Monte Carlo~\cite{Muser98,Hetenyi99,Hetenyi01}
and is known to be due to the ordering tendency of higher energy states (the
states with angular momentum zero are disordered as they are of polar
symmetry, the first odd angular momentum states are ordered).  In both cases
we have found the transition to be of second order.

\begin{figure}[htp]
\begin{center}
\vspace{1cm}
\includegraphics[width=7cm,height=5cm]{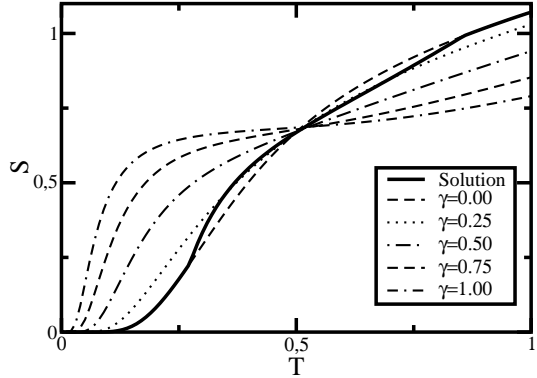}
\end{center}
\caption{Entropy calculations for a system without crystal field.  The
  comparison is for the entropy of the actual system (solution of the
  mean-field equations) and for fixed order parameter
  ($\gamma=0.00,0.25,0.50,0.75,1.00$).}
\label{fig:entropy_cf0.00}
\end{figure}

Calculations for the order parameter are presented in Fig. \ref{fig:op2d} for
a system with $X=2$ ($U_0=3.50$ reentrant region).  In the case of no crystal
field both transitions are manifestly second-order.  As the temperature is
increased the order parameter is zero until $T\approx 0.27$, it increases up
to $T\approx .5$, then the slope switches sign and decreases until $T\approx
.86$.  Subsequently the order parameter is zero.  The effect of the crystal
field is also shown in Fig.  \ref{fig:op2d}.  The order parameter for the
system with crystal field shows no discontinous change in the slope of the
order parameter, however a change in sign of the slope occurs at $T\approx
.5$ as in the case of no crystal field.  Another feature of the crystal field
is the appearance of a metastable state with negative order parameter as
shown in Fig. \ref{fig:op2d}.

The behaviour of the entropy for the system with $X=2$ without crystal field
is shown in Figure \ref{fig:entropy_cf0.00}.  As has been shown for the three
dimensional case ~\cite{Freiman05}, the entropy displays an anomaly in the
case of the reentrant phase diagram.  The entropy curves for the fixed order
parameter show qualitatively different behavior above and below $T=0.5$,
where the slope of the order parameter switches sign (Fig.  \ref{fig:op2d}).
The entropy of the disordered state ($\gamma=0$) is the lowest below the
temperature $T=0.5$, and the entropy increases as the system orders.  This
behaviour is unexpected from a classical point of view.  Above $T=0.5$ the
entropy of the ordered state is the lowest, and it increases upon
disordering, as expected based on the classical view.
\begin{figure}[htp]
\begin{center}
\vspace{1cm}
\includegraphics[width=7cm,height=5cm]{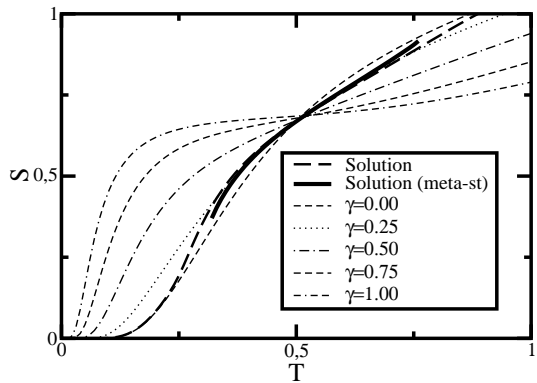}
\end{center}
\caption{Entropy calculations for a system with crystal field ($U_1=0.01$).
  The comparison is for the entropy of the actual system (solution of the
  mean-field equations), the meta-stable solution, and for fixed order
  parameter ($\gamma=0.00,0.25,0.50,0.75,1.00$).  The coupling constant is
  $U_0=3.5$}
\label{fig:entropy_cf0.01}
\end{figure}

This unusual feature can be understood from considering the expression of the
entropy for the quantum mechanical system $S=-\sum_i P_i\mbox{ln}P_i$, where
$P_i$ denote the probability for a particular state.  In the quantum
mechanical system the states are obtained after diagonalizing the Hamiltonian
(in the corresponding classical system the sum in the expression for the
entropy is an integral over the angles, and the probability is a function of
the angles as well).  As the lowest state, which dominates the behaviour of
the system at low temperatures (i.e. has the highest probability),
corresponds to a disordered state, it is not surprising that the entropy
decreases and that simultanously the system disorders.  In the state-space to
which the probabilities in the entropy expression refer the number of
possible states does in fact decrease (i.e. in that sense the system orders),
however the states themselves are disordered in real space. In some sense this
picture is similar to Bose-Einstein condensation~\cite{Leggett01}, where the
single lowest state becomes populated (and which corresponds to a state that
is spatially disordered), with the important difference that here the state
is not a collective state.
\begin{figure}[htp]
\begin{center}
\vspace{1cm}
\includegraphics[width=7cm,height=5cm]{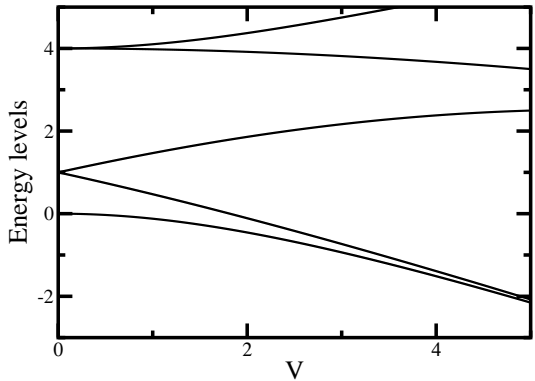}
\end{center}
\caption{Energy levels for a system with potential $V\mbox{cos}(2\phi)$.}
\label{fig:planar_levels}
\end{figure}

We also note that the entropy of the solution of the mean-field equations
corresponds to the disordered case below and above the phase transition
points.  At the phase transition points the slope of the entropy is
discontinuous.

The effect of the crystal field on the entropy is shown in Fig.
\ref{fig:entropy_cf0.01} ($U_1=0.01$).  The same behavior is observed with
regard to the ordering pattern as in Fig. \ref{fig:entropy_cf0.00}.  Below
the turning point of the slope of the order parameter (Fig. \ref{fig:op2d})
the entropy of the ordered state is higher than that of the disordered state.
Here the slope of the entropy does not change discontinously as a function of
temperature, as no phase transitions are experienced.
\begin{figure}[htp]
\begin{center}
\vspace{1cm}
\includegraphics[width=7cm,height=5cm]{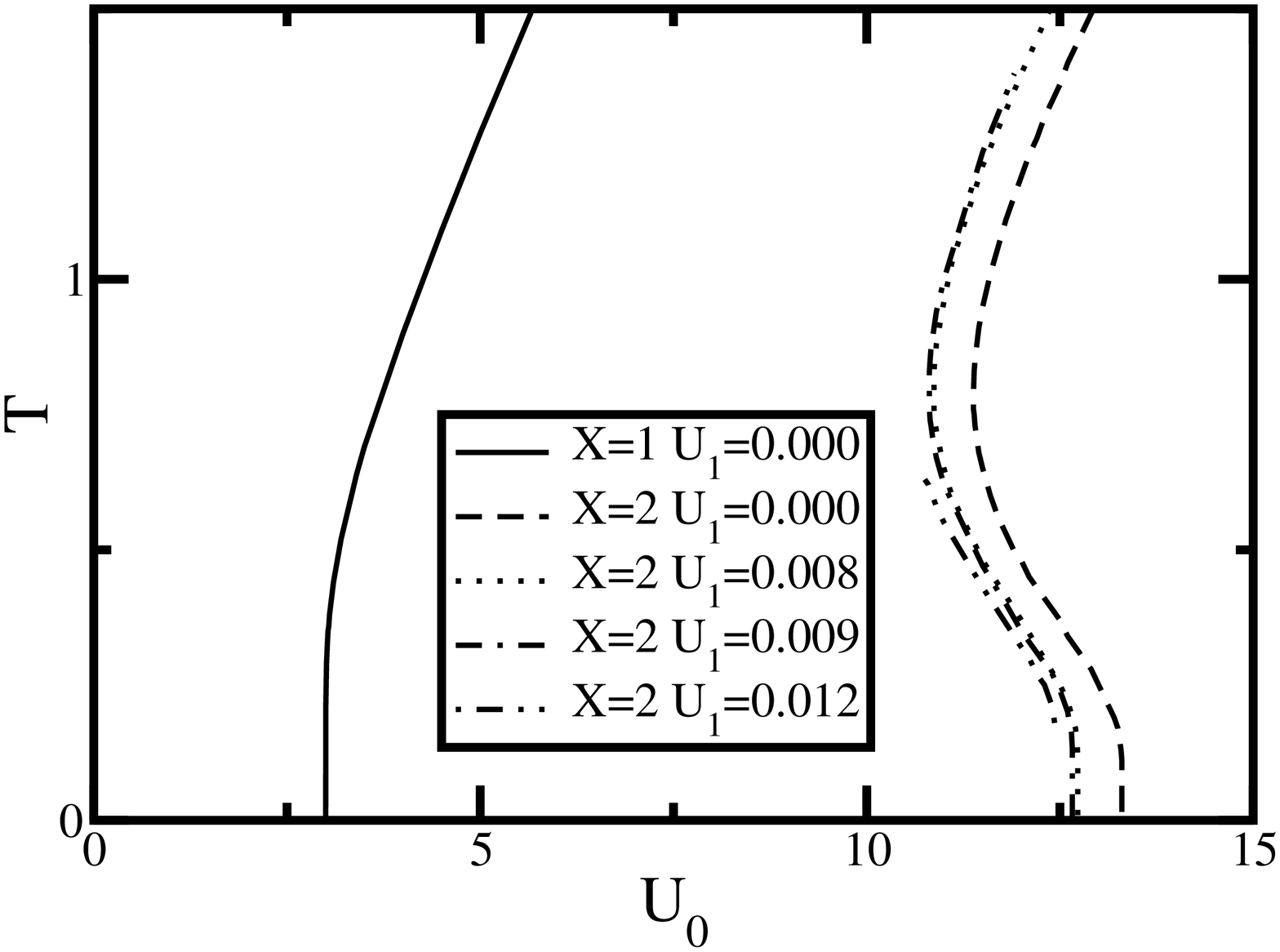}
\end{center}
\caption{Phase diagram for linear rotors, $X=1$ and $X=2$, with several
  values of the crystal field in the case of the latter.}
\label{fig:phase_diagram_3d}
\end{figure}

The entropy anomaly can also be understood in terms of the local energy
spectrum.  The eigenvalues of a planar rotor with potential
$V\mbox{cos}(2\phi)$ are shown in Fig. \ref{fig:planar_levels} as a function
of $V$.  At $V=0$ (disordered state) the ground state is a singlet and the
first excited state is doubly degenerate.  As $V$ is increased the degeneracy
of the first excited state is split, and the lower energy state becomes
degenerate with the ground state adding a factor of $R\mbox{ln}2$ to the
entropy at low temperatures.

\section{Coupled rotors in three dimensions}

\label{sec:3D}

In this section we calculate the mean-field phase diagram of a system of
coupled three-dimensional rotors under a crystal field.
\begin{equation}
\label{eqn:H_3D}
H = B \sum_{i=1}^N \hat{L}_i^2  - \frac{U}{2} \sum_{\langle
  i,j \rangle} \mbox{P}_{Xm}(\Omega_i)  \mbox{P}_{Xm}(\Omega_j) 
   - U_1 \sum_{i=1}^N \mbox{P}_{X0}(\Omega_i),
\nonumber
\end{equation}
where $\mbox{P}_{X0}(\Omega_i)=\sqrt{4\pi/(2X+1)}\mbox{Y}_{X0}(\Omega_i)$.
The mean-field approximation to the Hamiltonian in Eq. (\ref{eqn:H_3D})
results in
\begin{equation}
\label{eqn:H_MF_3D}
H_{MF} = B \sum_{i=1}^N \hat{L}_i^2
- (U_0\gamma + U_1) \sum_{i=1}^N   \mbox{P}_{X0}(\Omega_i)   
+ \frac{U_0 N}{2}\gamma^2.
\end{equation} 

The phase diagrams for the two cases $X=1$ and $X=2$, and for several crystal
fields in the case of the latter, are shown in Figure
\ref{fig:phase_diagram_3d}.  For the systems without crystal field, the two
features identified in the previous section in the case of the planar rotors,
namely the stronger ordering tendency in the $X=1$ case, and the reentrant
phase transition in the $X=2$ case are present in the case of linear rotors
as well.  An important difference is that the $X=2$ case exhibits a first
order phase transition.  An unusual feature develops upon turning on the
crystal field.  As shown before~\cite{Freiman03} the crystal field gives rise
to critical points which separate regions in the phase diagram where the
transition is first order from regions where no phase transition occurs,
rather a continuous increase in the order parameter (the exact quantitative
features of the phase diagram are explained in Ref. \cite{Freiman03}).
\begin{figure}[htp]
\begin{center}
\vspace{1cm}
\includegraphics[width=7cm,height=5cm]{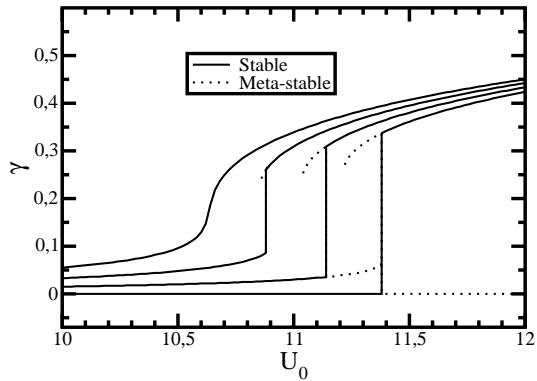}
\end{center}
\caption{Order parameter for different values of the crystal field as a
  function of the coupling constant $U_0$ at a temperature of $T=0.75$ The of
  the crystal field from left to right are $U_1=0.012,0.008,0.004,0.000$.
  The dotted lines indicate the value of the order parameter for
  metastable states.}
\label{fig:op_3d}
\end{figure}

In Figure \ref{fig:op_3d} we show the order parameter as a function of the
coupling constant at a temperature of $T=0.75$ (approximately where the
reentrant turning point occurs) for the $X=2$ system.  The calculations are
presented for different values of the crystal field
$U_1=0.012,0.008,0.004,0.000$.  The dotted lines indicate the meta-stable
states.  As usual in first-order phase transitions, as the parameter $U_0$ is
varied a meta-stable state develops before the phase transition, which
becomes the stable state upon crossing the phase transition point.
Simultaneously the stable state becomes meta-stable.  When no crystal field
is present we found that as $U_0$ is increased from the left, the ordered
meta-stable phase first appears at $U_0\approx 11.2$ and becomes the stable
state at $U_0=11.38$.  Subsequently the disordered phase $\gamma=0$ becomes
metastable.  As the crystal field is turned on the range where metastability
is encountered decreases.  For $U_1=0.004$, as $U_0$ is increased from the
left we find evidence for a meta-stable phase at $U_0 \approx 11.0$, the
phase transition is encountered at $U_0=11.14$, but the less ordered phase
(which was stable at $U_0\leq11.14$ persists as a meta-stable phase until
$U_0\approx11.4$.  For $U_1=0.008$ the phase transition is found at
$U_0=10.88$ and meta-stability is encountered only in a range $\approx.04$
around the phase-transition point.  For $U_1=0.012$ no phase transition is
encountered, only a continuous increase in the order parameter.
\begin{figure}[htp]
\begin{center}
\vspace{1cm}
\includegraphics[width=7cm,height=5cm]{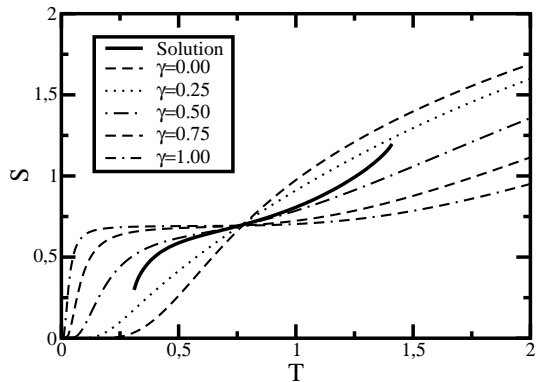}
\end{center}
\caption{Entropy of the ordered state and at fixed values of the order
  parameter for the system of linear rotors with $X=2$ at no crystal field,
  $U_0=12.50$.}
\label{fig:entropy_Xcr0.000}
\end{figure}

\begin{figure}[htp]
\begin{center}
\vspace{1cm}
\includegraphics[width=7cm,height=5cm]{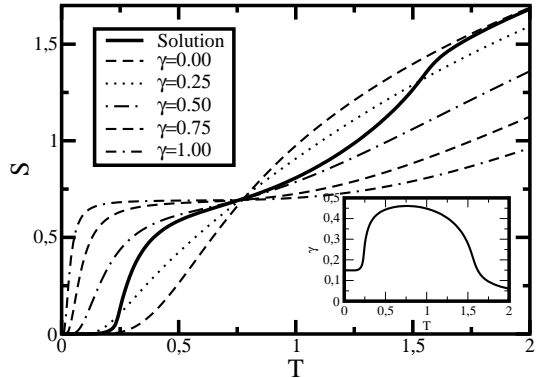}
\end{center}
\caption{Entropy of the ordered state and at fixed values of the order
  parameter for the system of linear rotors with $X=2$ at a crystal field of
  $U_1=0.018$, $U_0=12.00$.  The inset shows the order parameter. }
\label{fig:entropy_Xcr0.018}
\end{figure}

The entropy curves verify the general tendency shown in the case of linear
rotors in the previous section.  In Figures \ref{fig:entropy_Xcr0.000} and
\ref{fig:entropy_Xcr0.018} the value of the entropy corresponding to the
solution are shown as well as the value of the entropy at fixed order
parameter for the case without crystal field ($U_0=12.50$) and with a crystal
field of $U_1=0.018$ ($U_0=12.00$).  The inset shows the value of the order
parameter at $U_1=0.018$ as a function of temperature: as the temperature is
decreased the order parameter increases, it experiences a turning point at
$T=0.75$ and then begins to decrease.  This happens continuously, without any
phase transition.  The entropy of the ordered state, as was the case for the
planar rotors, is higher at low temperature ($T\leq 0.75$) than that of the
disordered state.  Thus the reversal of ordering as the temperature is cooled
appears to be correlated with the entropy anomaly, however, whether the
disordering occurs as a result of a phase transition is not.  In the absence
of the crystal field the quantum melting phase transition is second order for
planar rotors, first order for linear rotors.  When a crystal field is turned
on the phase transition is absent for planar rotors, whereas a more
complicated situation develops for linear rotors (see Figures
\ref{fig:phase_diagram_3d} and \ref{fig:op_3d} and Ref. \cite{Freiman05}),
but if the crystal field is large enough the ordering and disordering also
happens continuously.  The entropy anomaly accompanies all of these ordering
patterns.

\begin{figure}[htp]
\begin{center}
\vspace{1cm}
\includegraphics[width=7cm,height=5cm]{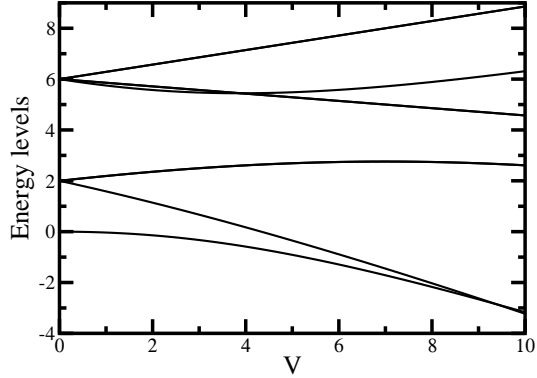}
\end{center}
\caption{Energy levels for a system with potential $VY_{20}(\Omega)$.}
\label{fig:linear_states}
\end{figure}

The energy levels for the linear rotors in an external potential of
$VY_{20}(\Omega)$ are shown in Fig. \ref{fig:linear_states}.  As in the case
of the planar rotors increase of $V$ from zero causes one state to move down
and approach the ground state causing an increase of $\approx Rln2$ in the
entropy.  

\section{Conclusions}
\label{sec:conclusions}

We have presented a comparative review of the mean-field theory of different
types of coupled rotors.  We have considered planar and linear rotors in
dipolar and quadrupolar potentials.  These models have corresponding physical
realizations: diatomic molecules (heteronuclear in the dipolar case,
homonuclear in the quadrupolar case) physisorbed on surfaces (two dimensional
system) or in the solid phase (three dimensional system).  The dipolar
potentials in both cases lead to a usual phase diagram where above a
particular value of the coupling constant the temperature vs. coupling
constant phase diagram increases with coupling constant.  The quadrupolar
potentials lead to reentrant phase diagrams in both cases: at low
temperatures, for some values of the coupling constant, quantum melting takes
place.  The phase transition for the planar rotors is always second order.
For the linear rotors the dipolar potential leads to a second-order phase
transition, in the quadrupolar potential the phase transition is first order.

We have also shown the different effects found when the systems are subjected
to a crystal field.  For the dipolar potentials the crystal field causes a
disappearance of the phase transition line, as temperature is decreased, and
as the coupling constant is increased only a continuous increase in the order
parameter is found.  As the ordering increases a metastable state is also
found with a negative order parameter.  We have also found this for the
planar rotors coupled via a quadrupolar potential.  For the linear rotors the
situation is more complicated.  As previously found\cite{Freiman05}
increasing the crystal field causes the appearance of critical points which
separate the phase diagram into lines where the phase transition is first
order from regions where no phase transition, but a continuous change in the
order parameter occurs.  An interesting accompanying feature is that where
there is a phase transition, the range in which a metastable state is found
decreases with the strength of the crystal field.

The reentrance in the case of the quadrupolar systems is accompanied by an
entropy anomaly: if the order parameter is held fixed the entropy of the
ordered state is higher at low temperatures than that of the disordered
state.  The situation reverses when the temperature is increased.  This
entropy anomaly is present in all the systems which exhibit quantum melting,
irrespective whether the melting takes place via a phase transition (either
first or second order), or via a continuous change in the order parameter.
Calculation of the spectrum of the mean-field potentials shows that the
entropy anomaly can be explained in terms of the change in the degeneracies
of states as a function of the coupling constant, as the ground state becomes
doubly degenerate.  It can also be argued that the entropy anomaly is a
natural consequence of quantum mechanics: the entropy decreases with
temperature, as a single state begins to dominate, but this single state is a
delocalized one (zero angular momentum state), hence it is disordered.

%%%%%%%%%%%%%%%%%%%%%%%%%%%%%%%%%%%%%%%%%%%%%%%%%%%%%%%%%%%
%\begin{acknowledgements}
%%%%%%%%%%%%%%%%%%%%%%%%%%%%%%%%%%%%%%%%%%%%%%%%%%%%%%%%%%%
%\end{acknowledgements}

%%%%%%%%%%%%%%%%%%%%%%%%%%%%%%%%%%%%%%%%%%%%%%%%%%%%%%%%%%%%%%%%%%%

\end{document}